\documentclass[journal]{IEEEtran}
\usepackage{graphicx,xcolor}
\usepackage[T1]{fontenc}
\usepackage{lmodern}
\usepackage{textcomp}
\usepackage{latexsym}
\usepackage{cite}
\usepackage{bm}

\usepackage{amsmath,amssymb,amsfonts}
\usepackage{algorithmic}
\usepackage{textcomp}
\usepackage{wrapfig}
\usepackage{unit}
\usepackage{stfloats}
\usepackage{multirow}
\usepackage[T1]{fontenc}
\usepackage{enumerate}
\usepackage{booktabs}
\usepackage{balance}
\usepackage{url}
\usepackage{threeparttable}
\usepackage{siunitx}
\usepackage{tabularx}
\usepackage{pifont}
\usepackage[super]{nth}
\usepackage{subcaption}
\usepackage{algorithm}

\newcommand{\TT}        {{\rm T}}

\ifCLASSINFOpdf
\else
\fi

\begin{document}
\title{Deformation-Aware Observation Modeling \\ for Radar-Based Human Sensing \\ via 3D Scan--Depth Sequence Fusion}
\author{Guangqi~Shi, Kimitaka~Sumi,\IEEEmembership{Graduate Student Member, IEEE}, and Takuya~Sakamoto, \IEEEmembership{Senior Member, IEEE}
\thanks{G.~Shi, K.~Sumi, and T.~Sakamoto are with the Department of Electrical Engineering, Graduate School of Engineering, Kyoto University, Kyoto, Kyoto 615-8510, Japan (e-mail: sakamoto.takuya.8n@kyoto-u.ac.jp).}}

\markboth{}%
         {Shi \emph{et al.}: Deformation-Aware Observation Modeling for Radar-Based Human Sensing via 3D Scan--Depth Sequence Fusion}


\maketitle
  \begin{abstract}
    Non-contact radar-based human sensing is often interpreted using simplified motion assumptions. However, respiration induces non-rigid surface deformation of the human body that impacts electromagnetic wave scattering and can degrade the robustness of measurements. To address this, we propose a surface-deformation-aware observation model for radar-based human sensing that fuses static high-resolution three-dimensional scanner measurements with temporal depth camera data to represent time-varying human surface geometry. Non-rigid registration using the coherent point drift algorithm is employed to align a static template with dynamic depth frames. Frame-wise electromagnetic scattering is subsequently computed using the physical optics approximation, allowing the reconstruction of intermediate-frequency radar signals that emulate radar observations. Validation against experimental radar data demonstrated that the proposed model exhibited greater robustness than a depth-sequence-only model under low-signal-quality conditions involving complex surface dynamics and multiple reflective sites. For two participants, the proposed model achieved higher Pearson correlation coefficients of 0.943 and 0.887 between model-derived and experimentally measured displacement waveforms, compared with 0.868 and 0.796 for the depth-sequence-only model. Furthermore, in a favorable case characterized by a single relatively-stationary reflective site, the proposed method achieved a correlation coefficient of 0.789 between model-derived and experimentally measured in-phase-quadrature magnitude variations. These results suggest that our sensor-fusion-based deformation-aware observation modeling can realistically reproduce radar observations and provide physically grounded insights into the interpretation of radar measurement variations.
\end{abstract}

\IEEEpeerreviewmaketitle


\section{Introduction}
\label{sec:introduction}

\IEEEPARstart{P}{hysiological} indicators such as heart rate, respiratory rate, and blood pressure play important roles in both medical monitoring and daily health management. Previous studies in hospitalized patients on general wards have shown that abnormalities in these vital signs can be significant predictors of adverse clinical events including cardiac arrest and sudden deterioration~\cite{DeVita2010consensus,Ludikhuize2012deterioration,Ludikhuize2014rrs}. Although contact-based sensors such as electrocardiographs are widely used in routine care, they cannot be applied to patients with skin allergies or burn injuries~\cite{Singh2021review}, and their use may be impractical in scenarios requiring long-term and unobtrusive monitoring such as neonatal care, elderly home monitoring, or sleep monitoring~\cite{Lin2017sleep,Pereira2019infant}.

Radar-based non-contact monitoring has emerged as a promising alternative to contact-based sensors, enabling remote physiological sensing with minimal intrusion on the monitored individual~\cite{Paterniani2023overview,Kouhalvandi2025survey}. Nevertheless, it should be noted that despite ongoing technological progress, degradation of the signal quality caused by body posture and orientation persists~\cite{Nahar2018model}, and this degradation has not yet been elucidated with a physically-grounded electromagnetism-based measurement interpretation.

A key challenge stems from the non-rigid nature of human anatomy. During respiration, the torso undergoes continuous deformation, particularly in the thoracic and abdominal regions. These dynamic surface changes alter the electromagnetic scattering characteristics observed by radar, complicating both the interpretation of measurements and the simulation of radar signals. Accurate modeling of such time-varying scattering behavior therefore necessitates an observation modeling framework capable of capturing the non-rigid geometric changes of the human body in motion.

Several studies have attempted to interpret radar observations through electromagnetic simulations. Quaiyum~\emph{et al.} employed a multilevel fast multipole algorithm (MLFMA) to model vital sign detection using anatomically simplified body representations~\cite{Quaiyum2018modeling}. Mukherjee~\emph{et al.} integrated synthetic displacement models representing cardiac and respiratory motion into a human model and applied the physical optics (PO) approximation to estimate time-varying scattering behavior~\cite{Mukherjee2024animate}. In another approach, Koshisaka and Sakamoto utilized static 3D human models to identify radar reflective sites with PO-based analysis~\cite{Koshisaka2025pulse}. However, these prior works relied on either idealized models or static surfaces, and none of them directly modeled the time-resolved scattering characteristics of a real human subject undergoing natural respiratory motion.

Depth cameras offer a means to capture dynamic surface geometry in real time. However, their limited sensing accuracy and susceptibility to measurement noise compromise the fidelity of the acquired data~\cite{Izadi2011kinectfusion, Zhang2012kinect}, making them unsuitable for accurate electromagnetic simulation of radar measurements. In contrast, 3D scanners produce high-fidelity point cloud data under static conditions and are well suited for surface reconstruction~\cite{Jeught2016profilometry,Berger2017recon}, as well as subsequent electromagnetic simulation, but are inherently incapable of capturing transient respiratory dynamics.

To overcome these limitations, this study introduces a deformation-aware observation modeling approach for radar-based human sensing that integrates time-series depth camera data with high-resolution 3D scanner data. Robust non-rigid registration techniques---an area that has seen significant progress in recent years~\cite{Huang2021registration}---are essential for aligning these complementary datasets. By appropriately integrating the temporal resolution of the depth camera and the geometric fidelity of the 3D scanner, it is possible to reconstruct respiratory surface deformation over time and estimate the associated electromagnetic scattering using the PO approximation. This approach seeks to bridge the gap between empirical radar observations and measurement-oriented modeling, and can further reveal how dynamic surface geometry alters radar scattering in non-contact physiological monitoring.

\section{Electromagnetic Scattering Analysis and Radar Signal Reconstruction}
\label{sec2}
\subsection{Electromagnetic Scattering Analysis}
\label{subsec2a}
In attempts to identify dominant scattering regions on the surface of the human body, the PO approximation is commonly adopted to simulate electromagnetic scattering. In such procedures, the human body surface is modeled as a perfect electric conductor and the current distribution is approximated using the incident magnetic field and the surface normal vectors
\cite{Koshisaka2025pulse,Sakamoto2018adaptive,Konishi2019automatic,sumi}.

In this formulation, the approximate surface current density $\bm{J}(\bm{r}_\mathrm{S})$ at each point $\bm{r}_\mathrm{S} \in S$ on the surface $S$ is given by
\begin{equation}
    \bm{J}(\bm{r}_\mathrm{S}) = 2\hat{\bm{n}}(\bm{r}_\mathrm{S}) \times \bm{H}^\mathrm{inc}(\bm{r}_\mathrm{S}) ,
\label{eq:PO-approximation}
\end{equation}
where $\hat{\bm{n}}(\bm{r}_\mathrm{S})$ denotes the unit surface normal at $\bm{r}_\mathrm{S}$, oriented outward from the illuminated surface, i.e., in the direction opposite to the incident wave propagation.

Assuming that the transmitting antenna is modeled as an infinitesimal dipole oriented along the $z$-axis and located at the origin, the incident magnetic field at position $\bm{r}_\mathrm{S}$ is expressed as
\begin{equation}
    \bm{H}^\mathrm{inc}(\bm{r}_\mathrm{S}) = \left[-H_\phi(\bm{r}_\mathrm{S})\sin{\phi},\,H_\phi(\bm{r}_\mathrm{S})\cos{\phi},\,0\right]^\TT ,
\label{eq:H_inc(r)}
\end{equation}
where $\phi$ denotes the azimuth angle.

Given the surface current density $\bm{J}(\bm{r}_\mathrm{S})$, the electric field $\bm{E}(\bm{r})$ at an observation point $\bm{r}$ is expressed using the dyadic Green's function $\bar{\bm{G}}(\bm{r};\bm{r}_\mathrm{S})$ as
\begin{align}
    \bm{E}(\bm{r}) &= -\jj \omega \mu \iint_S \bar{\bm{G}}(\bm{r};\bm{r}_\mathrm{S}) \cdot \bm{J}(\bm{r}_\mathrm{S}) \, \dd S , \\
     \bar{\bm{G}}(\bm{r};\bm{r}_\mathrm{S}) &= \left( \bar{I} + \frac{1}{k_0^2} \nabla\nabla \right)\frac{\mathrm{e}^{-\jj k_0 R}}{4 \pi R},
\end{align}
where $\omega$ denotes the angular frequency, $\mu$ is the magnetic permeability, $k_0$ is the free-space wavenumber, $R = \|\bm{r} - \bm{r}_\mathrm{S}\|$, $\bar{I}$ is the unit dyad, and $\nabla\nabla$ represents the dyadic differential operator with respect to the observation coordinate $\vct{r}$ \cite{Sarabandi}.

This formulation enables the computation of the electric field at a given observation point. However, it does not indicate which specific regions on the body surface contribute significantly to the received electric field. To address this issue, Shijo \emph{et al.}~\cite{shijo2004visualization} proposed a visualization technique for identifying high-frequency diffraction contributors using a weighting function referred to as the eye function.

The eye function $w_\mathrm{eye}(\bm{r}_\mathrm{S};\bm{r}_0)$, defined in the vicinity of a surface point $\bm{r}_0 \in S$, is given by
\begin{equation}
  \resizebox{\linewidth}{!}{$
    w_\mathrm{eye}(\bm{r}_\mathrm{S};\bm{r}_0) =
    \begin{cases}
        \dfrac{1}{2} \left\{\cos{\left(\dfrac{\pi\|\bm{r}_\mathrm{S}-\bm{r}_0\|}{a_0}\right)} + 1\right\} & (\|\bm{r}_\mathrm{S}-\bm{r}_0\|\leq a_0) \\
        0 & (\|\bm{r}_\mathrm{S}-\bm{r}_0\|> a_0)
    \end{cases} ,
  $}
\label{eq:eye_func}
\end{equation}
where $a_0$ denotes the radius of the weighting function.

The electric field intensity observed at $\bm{r}$ due to scattering from the vicinity of a surface point $\bm{r}_0$ is computed as
\begin{equation}
  \resizebox{\linewidth}{!}{$
    \left|\bm{E}_\mathrm{scat}(\bm{r};\bm{r}_0)\right| = \left|-\jj \omega \mu \iint_S w_\mathrm{eye}(\bm{r}_\mathrm{S};\bm{r}_0) \, \left(\bar{\bm{G}}(\bm{r};\bm{r}_\mathrm{S}) \cdot \bm{J}(\bm{r}_\mathrm{S})\right) \, \dd S \right| .
  $}
\label{eq:E_scat_mag}
\end{equation}
Repeating this procedure across all surface points allows the spatial distribution of scattering power over the body surface to be obtained.

\subsection{Radar Signal Reconstruction}
\label{subsec2b}
In a frequency-modulated continuous-wave (FMCW) radar system equipped with a linear virtual array consisting of $N_{\mathrm{R}}$ elements, the intermediate frequency (IF) signal corresponding to the $i$th virtual array element is expressed as
\begin{equation}
  \resizebox{\linewidth}{!}{$
    s_{\mathrm{IF},i}(\tau, t) = \sum_{k=1}^{K} A_{\mathrm{IF},i,k} \exp \left\{ \jj 4\pi \left( \frac{\gamma R_{i,k}(t)}{c} \tau + \frac{f_\mathrm{min} R_{i,k}(t)}{c}\right)\right\} ,
  $}
  \label{eq:IF}
\end{equation}
where $c$ is the speed of light, $\tau$ is the fast time, $\gamma = B/T_\mathrm{c}$ is the chirp rate defined as the ratio of the bandwidth $B$ to the chirp duration $T_\mathrm{c}$, $f_\mathrm{min}$ represents the starting frequency of the chirp, and $K$ is the number of scattering points. The term $R_{i,k}(t)$ denotes the distance between the $i$th virtual array element and the $k$th scattering point as a function of slow time $t$. The amplitude term $A_{\mathrm{IF},i,k}$ represents the magnitude of the IF signal component associated with the $i$th virtual array element and the $k$th scattering point.

Following \cite{sumi}, the scattering contribution is modeled using a complex scattering coefficient $|\bm{E}_\mathrm{scat}(\bm{r}_i;\bm{r}_k)|\eta_{i,k}$ for IF signal reconstruction. Here, $|\bm{E}_\mathrm{scat}(\bm{r}_i;\bm{r}_k)|$ denotes the field intensity computed using Eq.~\eqref{eq:E_scat_mag}, whereas $\eta_{i,k}$ represents the effective phase term.

To clarify how scattering points and their amplitudes are handled in~\cite{sumi}, we briefly summarize the formulation here. Let the true human body surface at time $t$ be represented as a smooth manifold $\mathcal{S}(t) \subset \mathbb{R}^3$. For each $t$, the depth camera observes a point cloud $\mathcal{P}^{\mathcal{S}}_{\mathrm{cam}}(t) = \big\{ \bm{p}^{\mathcal{S}}_{\mathrm{cam},k}(t) \big\}_{k}$, which samples the surface $\mathcal{S}(t)$. To mitigate the spatial random error inherent in per-frame depth data, a time-averaged geometry is constructed by averaging each sampled point over the observation interval $[0,T]$,
\begin{equation}
    \bar{\bm{p}}^{\mathcal{S}}_{\mathrm{cam},k}
    = \frac{1}{T}\int_0^T \bm{p}^{\mathcal{S}}_{\mathrm{cam},k}(t)\,\dd t,
    \quad
    \bar{\bm{p}}^{\mathcal{S}}_{\mathrm{cam},k} \in
    \bar{\mathcal{P}}^{\mathcal{S}}_{\mathrm{cam}},
\end{equation}
after which PO-based scattering analysis is performed once on this time-averaged point set $\bar{\mathcal{P}}^{\mathcal{S}}_{\mathrm{cam}}$.

For a fixed radar antenna position $\bm{r}_i$, the intensity of the scattering field $\big| \bm{E}_{\mathrm{scat}}(\bm{r}_i; \bar{\bm{p}}^{\mathcal{S}}_{\mathrm{cam},k}) \big|^2$ is evaluated across
$\bar{\mathcal{P}}^{\mathcal{S}}_{\mathrm{cam}}$. The points $\bar{\bm{p}}_k \in \bar{\mathcal{P}}^{\mathcal{S}}_{\mathrm{cam}}$ that exhibit locally dominant intensities exceeding a threshold $\theta_{\mathrm{scat}}$ are extracted as scattering centers. For each $\bar{\bm{p}}_k$, the corresponding distance $R_{i,k}(t)$ in Eq.~\eqref{eq:IF} is then obtained by selecting, at each time $t$, the point
$\bm{p}_k^{*}(t) \in \mathcal{P}^{\mathcal{S}}_{\mathrm{cam}}(t)$ whose direction with respect to the radar antennas is most closely aligned with that of $\bar{\bm{p}}_k$.

In addition, the scattered field magnitudes are assumed to be time-invariant:
\begin{equation}
  \big| \bm{E}_{\mathrm{scat}}(\bm{r}_i; \bm{p}^*_k(t)) \big|
  = \big| \bm{E}_{\mathrm{scat}}(\bm{r}_i; \bar{\bm{p}}_k) \big|,
  \label{eq:sumi_assumption_compact}
\end{equation}
such that, under the assumption of small respiratory-induced displacements relative to the propagation distance, the temporal variation of the scattered field is dominated by phase modulation through the path length $R_{i,k}(t)$. Accordingly, the model in~\cite{sumi} relies on two key assumptions:
(i) the dominant scattering centers remain approximately on the same lines of sight with respect to the radar antennas throughout the respiratory motion, and (ii) the corresponding scattering field intensities remain constant over time.

In contrast, Mukherjee \emph{et al.} used a 3D human phantom with synthetically imposed cardiac and respiratory motion and performed PO-based scattering analysis on temporally sampled snapshots (e.g., every $50\,\mathrm{ms}$)
~\cite{Mukherjee2024animate}.
In their approach, the received signal is reconstructed using per-frame scattering results, thereby explicitly accounting for temporal variations in both surface geometry and scattering strength. Table~\ref{tab:sim_framework_comparison} summarizes these PO-based frameworks in terms of the human model adopted and the number of PO evaluations required to synthesize a 60-s radar signal.

\begin{table}[tb]
  \centering
  \caption{Comparison of PO simulation frameworks for radar-based vital sign monitoring.}
  \label{tab:sim_framework_comparison}
  \small
  \renewcommand{\arraystretch}{1.2}
  \resizebox{\linewidth}{!}{%
  \begin{tabular}{@{}l l c@{}}
    \toprule
    \textbf{Study} &
    \textbf{Human model} &
    \shortstack{\textbf{PO}\\\textbf{evaluations/60 s}} \\
    \midrule
    Mukherjee \emph{et al.}~\cite{Mukherjee2024animate}
      & Animated human phantom
      & 1200 \\

    Sumi \& Sakamoto~\cite{sumi}
      & Depth camera point clouds
      & 1 \\

    Proposed method
      & Scanner--camera integration
      & 900 \\
    \bottomrule
  \end{tabular}%
}
\end{table}

\section{Sensor-Driven Modeling for Radar Signal Analysis}
\label{sec3}
\subsection{Integration of 3D Scanner and Depth Camera Data}
\label{subsec3a}
In this study, we constructed a dynamic model that accommodates temporal variations by iteratively fitting a high-resolution point cloud template $\mathcal{P}^{\mathcal{S}}_{\mathrm{scan}}$ obtained from a 3D scanner to the frame-by-frame point cloud data
$\mathcal{P}^{\mathcal{S}}_{\mathrm{cam}}(t)$
captured by a depth camera. This data integration method enables the construction of a time-varying deformation model of the human body surface while mitigating the influence of random errors arising from the limited depth accuracy of the depth camera.

To address the absence of deterministic point-wise correspondence between the 3D-scanner-derived template $\mathcal{P}^{\mathcal{S}}_{\mathrm{scan}}$ and the point clouds $\mathcal{P}^{\mathcal{S}}_{\mathrm{cam}}(t)$ captured in each depth frame, we employ the coherent point drift (CPD) algorithm~\cite{Myronenko2010cpd}, in which non-rigid point set alignment is formulated as a probability density estimation problem.

In the CPD algorithm, the source point cloud (i.e., the template) $Y \subset \mathcal{P}^{\mathcal{S}}_{\mathrm{scan}}$, consisting of $M$ points $\{\bm{y}_m\}_{m=1}^M$, is modeled as a Gaussian mixture model (GMM), where each $\bm{y}_m$ serves as the centroid of a $D$-dimensional Gaussian distribution (typically $D=3$ for 3D point clouds). In contrast, the target point cloud $X \subset \mathcal{P}^{\mathcal{S}}_{\mathrm{cam}}(t_0)$ containing $N$ points $\{\bm{x}_n\}_{n=1}^N$ from a single depth frame at $t=t_0$ is treated as observed data sampled from an underlying latent probability distribution corresponding to the true human body surface $\mathcal{S}(t_0)$.

This formulation enables application of the expectation--maximization (EM) algorithm; in the expectation step, the expected likelihood that the target points are generated by the GMM is computed, while in the maximization step, the model parameters---namely, the centroid positions and the variance---are iteratively updated to maximize this expected likelihood. A regularization term is incorporated into the update process to enforce motion coherence among neighboring centroids, thereby ensuring that nearby regions of the template deform in a smooth and consistent manner. The basic flow of the CPD algorithm, originally introduced in~\cite{Myronenko2010cpd}, is summarized in Algorithm~\ref{alg:cpd} within the standard EM framework.

\begin{algorithm}[tb]
\renewcommand{\algorithmicrequire}{\textbf{Initialize:}}
\renewcommand{\algorithmicensure}{\textbf{Output:}}
\caption{Coherent point drift algorithm~\cite{Myronenko2010cpd}}
\label{alg:cpd}
\begin{algorithmic}[1]
\REQUIRE $W \gets \bm{0},\;\sigma^2 \gets \frac{1}{DMN}\sum_{m,n} \|\bm{x}_n - \bm{y}_m\|^2$

\REPEAT
    \STATE \textbf{E-step:}
    \STATE \quad Compute correspondence probabilities $P = [p_{mn}]$
    \[p_{mn} =
        \frac{\exp(-d_{mn}/2\sigma^2)}{\sum_{k=1}^M\exp(-d_{k,n}/2\sigma^2)+\frac{Mw}{N(1-w)} \left(2\pi\sigma^2 \right)^{{D}/{2}}}\]
    \[d_{mn}=\left\| \bm{x}_n-\left(\bm{y}_m + (G_{m:}W)^\top \right)\right\|^2\]
    \STATE \textbf{M-step:}
    \STATE \quad $T \gets Y + GW$
    \STATE \quad $W \gets (G+\lambda\sigma^2\tilde{P}^{-1})^{-1}(\tilde{P}^{-1}PX - Y)$
    \STATE
    \parbox[t]{0.95\linewidth}{$
    \begin{aligned}
    \quad \sigma^2 \gets \frac{1}{N_P D}\bigl\{
    &\mathrm{tr}(X^\top \breve{P} X)- 2\,\mathrm{tr}((PX)^\top T) \\
    &+ \mathrm{tr}(T^\top \tilde{P}T)\bigr\}
    \end{aligned}
    $}
    \STATE \textbf{Definitions:}
    \STATE \quad $G_{ij} = \exp (-\|\bm{y}_i - \bm{y}_j\|^2 / 2\beta^2)$
    \STATE \quad $\tilde{P}_{mm}=\sum_n p_{mn}$,\;
    $\breve{P}_{nn}=\sum_m p_{mn}$,
    \STATE \quad $N_P=\sum_{mn} p_{mn}$.
\UNTIL{convergence}
\end{algorithmic}
\end{algorithm}

\begin{figure}[tb]
    \centering
    \begin{minipage}[b]{0.2\textwidth}
        \centering
        \includegraphics[width=0.9\linewidth]{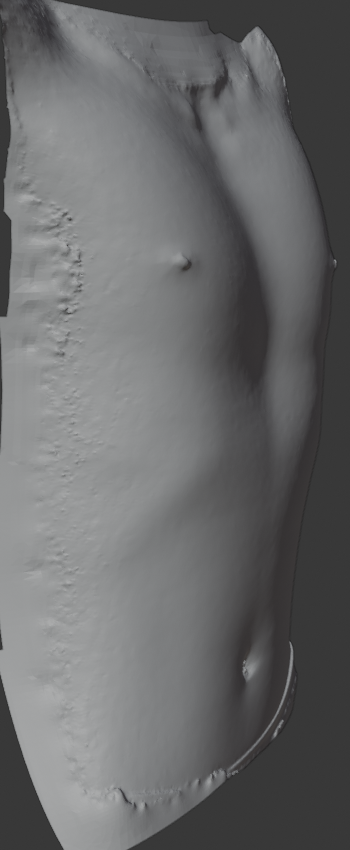}
        \caption{Example of the reconstructed surface for an end-expiratory-phase frame.}
        \label{fig:end_exp}
    \end{minipage}
    \hfill
    \begin{minipage}[b]{0.2\textwidth}
        \centering
        \includegraphics[width=0.9\linewidth]{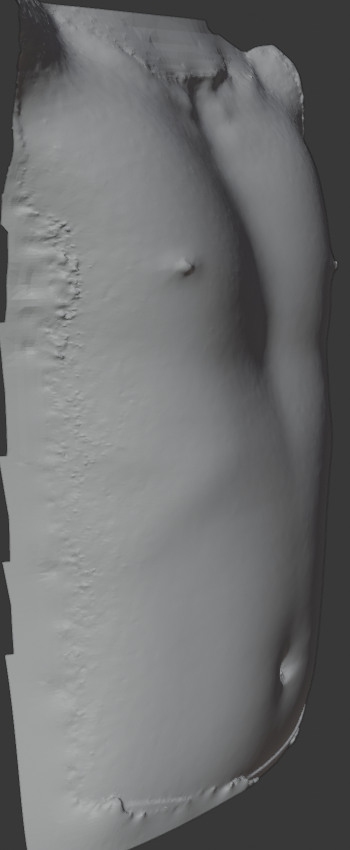}
        \caption{Example of the reconstructed surface for an end-inspiratory-phase frame.}
        \label{fig:end_insp}
    \end{minipage}
\end{figure}

\begin{figure}[tb]
    \centering
    \begin{minipage}[t]{0.2\textwidth}
        \centering
        \includegraphics[width=0.9\linewidth]{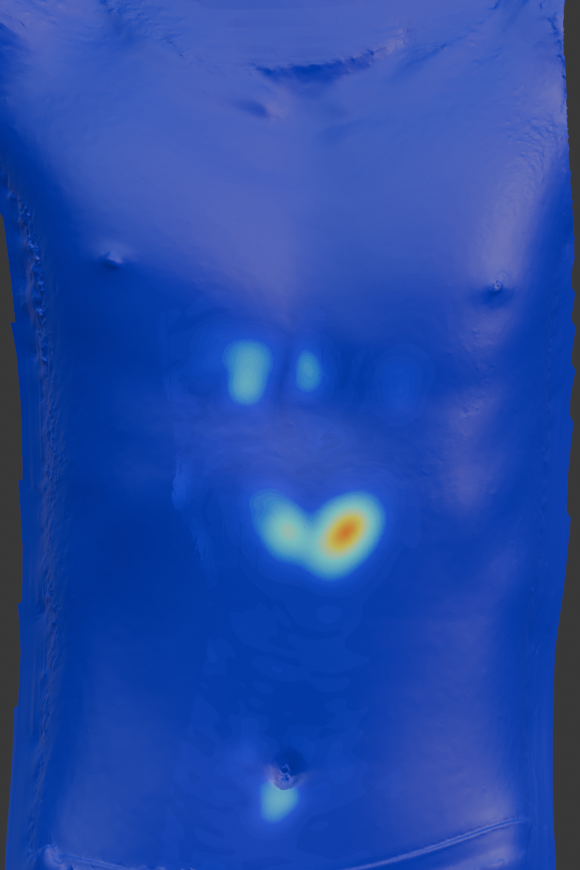}
        \caption{Example of the scattering power distribution for an end-expiratory-phase frame.}
        \label{fig:end_exp_scat}
    \end{minipage}
    \hfill
    \begin{minipage}[t]{0.2\textwidth}
        \centering
        \includegraphics[width=0.9\linewidth]{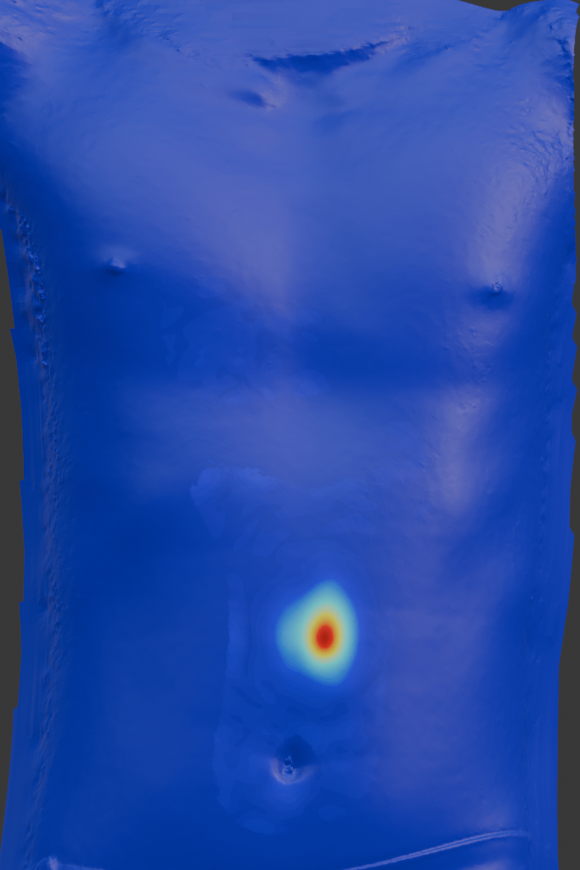}
        \caption{Example of the scattering power distribution for an end-inspiratory-phase frame.}
        \label{fig:end_insp_scat}
    \end{minipage}
\end{figure}

Figs.~\ref{fig:end_exp} and \ref{fig:end_insp} show representative examples of reconstructed surfaces obtained using the framework described above, while the corresponding estimated scattering field distributions are presented in Figs.~\ref{fig:end_exp_scat} and \ref{fig:end_insp_scat}.
The point cloud datasets were acquired using a Scantech iReal 2E 3D scanner (Scantech Co., Ltd., Hangzhou, China) and a Microsoft Azure Kinect DK depth camera (Microsoft Corp., Redmond, WA, USA). The specifications and operating parameters of the devices are summarized in Tables~\ref{tab:3D-scanner-spec} and \ref{tab:depth-camera-spec}, respectively. Specifically, Figs.~\ref{fig:end_exp} and \ref{fig:end_exp_scat} correspond to a frame captured at end-expiration (i.e., the onset of inspiration), whereas Figs.~\ref{fig:end_insp} and \ref{fig:end_insp_scat} correspond to a frame captured at end-inspiration (i.e., the onset of expiration).

\begin{table}[tb]
  \centering
  \caption{SPECIFICATIONS OF THE 3D SCANNER (IREAL 2E) USED IN THIS STUDY}
  \label{tab:3D-scanner-spec}
  \begin{tabular}{cc}
    \toprule
    \textbf{Specification} & \textbf{Details} \\
    \midrule
    Light source & Infrared VCSEL \\
    Measurement accuracy & $0.1\,\mathrm{mm}$ \\
    Scanner resolution (point distance) & $0.2 \sim 3\,\mathrm{mm}$ \\
    \bottomrule
  \end{tabular}
\end{table}

\begin{table}[tb]
  \centering
  \caption{SPECIFICATIONS OF THE DEPTH CAMERA (AZURE KINECT DK) USED IN THIS STUDY}
  \label{tab:depth-camera-spec}
  \begin{tabular}{cc}
    \toprule
    \textbf{Parameter} & \textbf{Value} \\
    \midrule
    Resolution & $1024 \times 1024$ \\
    Frames per second & $15$ \\
    Operating range & $0.25\,\mathrm{m}\sim 2.21\,\mathrm{m}$ \\
    Field of view (H $\times$ V) & $120^\circ \times 120^\circ$ \\
    \bottomrule
  \end{tabular}
\end{table}

\subsection{Proposed Method for Radar Signal Reconstruction}
\label{subsec3b}
A key advantage of fitting a static template to the time-series depth frames during data integration is the establishment of temporal consistency. Specifically, each point in the CPD-fitted scanner point cloud
\begin{equation}
  \hat{\mathcal{P}}^{\mathcal{S}}_{\mathrm{scan}}(t) = \mathcal{T}_t(\mathcal{P}^{\mathcal{S}}_{\mathrm{scan}};\mathcal{P}^{\mathcal{S}}_{\mathrm{cam}}(t))
\end{equation}
corresponds to the same local region of the human body surface across all frames. This is because the transformation $\mathcal{T}_t$ maps points in the static template $\mathcal{P}^{\mathcal{S}}_{\mathrm{scan}}$ to their estimated positions at time $t$, conditioned on the depth camera observation $\mathcal{P}^{\mathcal{S}}_{\mathrm{cam}}(t)$.

The resulting point cloud $\hat{\mathcal{P}}^{\mathcal{S}}_{\mathrm{scan}}(t)$ lies on the CPD-estimated evolving surface $\hat{\mathcal{S}}(t) \subset \mathbb{R}^3$, which serves as an approximation of the true time-varying surface $\mathcal{S}(t)$ while preserving the point-wise indexing of the static template $\mathcal{P}^{\mathcal{S}}_{\mathrm{scan}}$.
By contrast, when the depth camera alone is used, temporal consistency is limited to fixed observation rays
inherent to the depth image pixel grid, and point-wise correspondence across time is not guaranteed.

The proposed framework therefore enables the definition of a fixed index set $\mathfrak{K}$, such that for each $k \in \mathfrak{K}$, the sequence $\{\bm{p}_k(t)\}_t$ represents the time-varying trajectory of a single consistently-tracked surface point in $\hat{\mathcal{P}}^{\mathcal{S}}_{\mathrm{scan}}(t)$. In this study, $\mathfrak{K}$ is defined as
\begin{equation}
    \mathfrak{K} = \left\{ k \;\bigg|\; \exists\, t \text{ s.t. }
    \frac{\left| \bm{E}_{\mathrm{scat}}(\bm{r}_i; \bm{p}_k(t)) \right|^2}
    {\displaystyle\max_{t',\, \bm{p}_k'} \left| \bm{E}_{\mathrm{scat}}(\bm{r}_i; \bm{p}_k'(t')) \right|^2}
    \geq \theta_\mathrm{thresh} \right\},
\end{equation}
that is, the set of surface points that exhibit sufficiently strong scattering at least once during the observation period.

Because the indexing of $\hat{\mathcal{P}}^{\mathcal{S}}_{\mathrm{scan}}$ is invariant with respect to time, the scattering field magnitudes $|\bm{E}_\mathrm{scat}(\bm{r}_i;\bm{p}_k)|$ are naturally defined as time series. This leads to the following time-dependent formulation of the intermediate-frequency signal
\begin{equation}
  \resizebox{\linewidth}{!}{$
    s_{\mathrm{IF},i}(\tau, t) = \sum_{k=1}^{K} |\bm{E}_\mathrm{scat}(\bm{r}_i;\bm{p}_k;t)|\eta_{i,k} \exp \left\{ \jj 4\pi \left( \frac{\gamma R_{i,k}(t)}{c} \tau + \frac{f_\mathrm{min} R_{i,k}(t)}{c}\right)\right\}.
  $}
  \label{eq:IF_t}
\end{equation}

Although the formulation in Eq.~\eqref{eq:IF}, which was originally introduced in \cite{sumi}, is also time-varying through its dependence on $R_{i,k}(t)$, the proposed method introduces two important refinements. First, $\hat{\mathcal{P}}^{\mathcal{S}}_{\mathrm{scan}}(t)$ provides a more physically-grounded approximation of the true time-varying surface $\mathcal{S}(t)$ than does the time-averaged surface $\bar{\mathcal{P}}^{\mathcal{S}}_{\mathrm{cam}}$. This is because it preserves the temporal deformation of the surface geometry while suppressing depth-camera-specific artifacts via probabilistic non-rigid registration of a high-fidelity scanner template. Second, rather than relying on scattering centers derived from a static or time-averaged surface, the proposed approach employs a dynamically deforming surface model, enabling consistent computation of both the amplitude and phase of the radar echo at each time step.

\section{Experimental Performance Evaluation}
\label{sec4}
\subsection{Experimental Setup}
\label{subsec4a}
We conducted experiments to validate the proposed modeling framework and evaluate its performance. The experimental setup consisted of a Scantech iReal 2E 3D scanner (Scantech Co., Ltd., Hangzhou, China), a Microsoft Azure Kinect DK depth camera (Microsoft Corp., Redmond, WA, USA), and a millimeter-wave FMCW radar system operating at a center frequency of $79\,\mathrm{GHz}$ (T14RE\_01080108\_2D, S-Takaya Electronics Industry Co., Ltd., Okayama, Japan). The specifications of these devices are summarized in Tables~\ref{tab:3D-scanner-spec}, \ref{tab:depth-camera-spec}, and \ref{tab:radar-spec}, respectively.

\begin{figure}[tb]
        \centering
        \begin{minipage}{0.99\linewidth}
            \centering
            \includegraphics[width=0.7\linewidth]{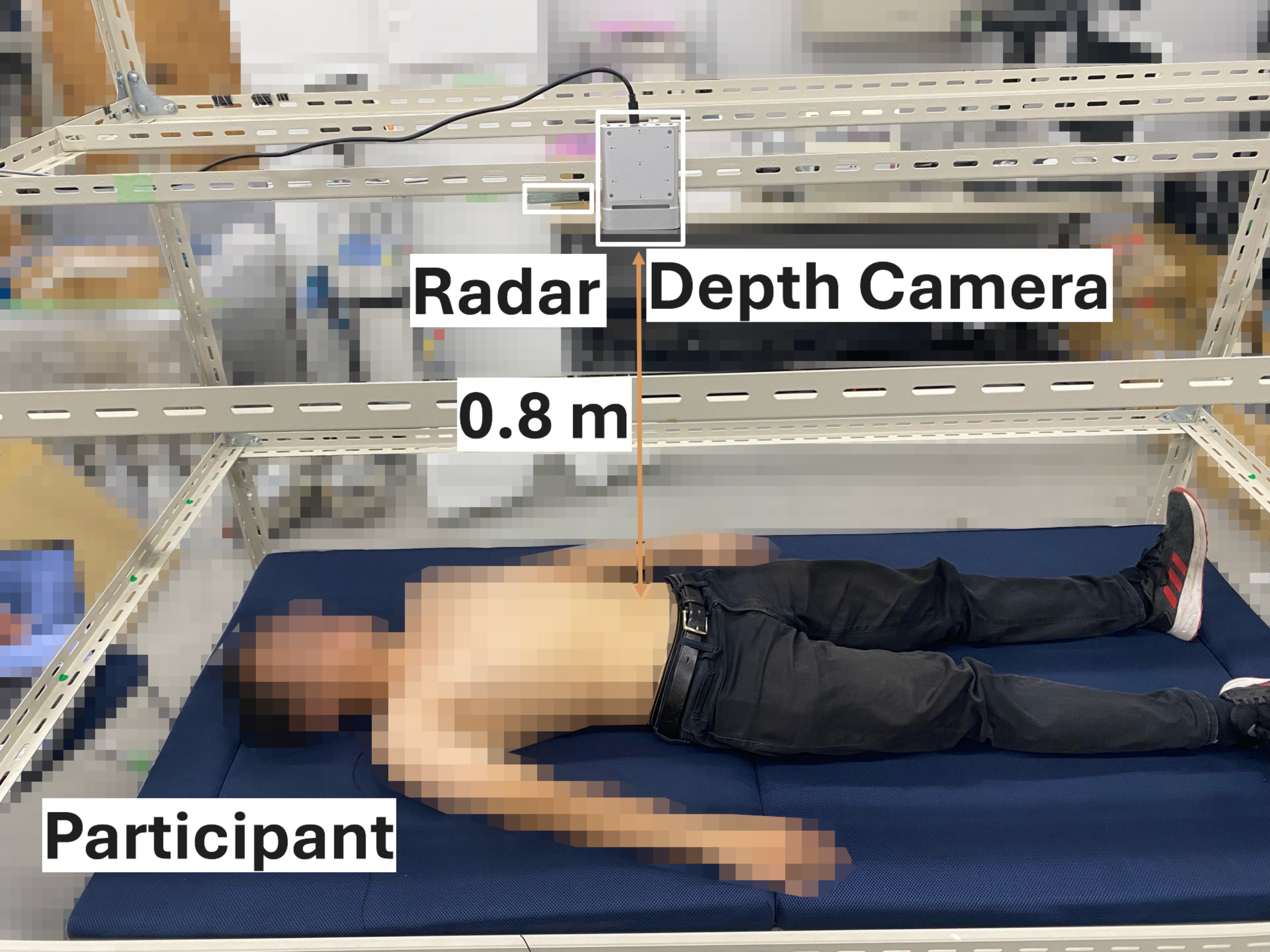}
            \subcaption{} \label{fig:scene}
        \end{minipage}
        \begin{minipage}{0.99\linewidth}
            \centering
            \includegraphics[width=\linewidth]{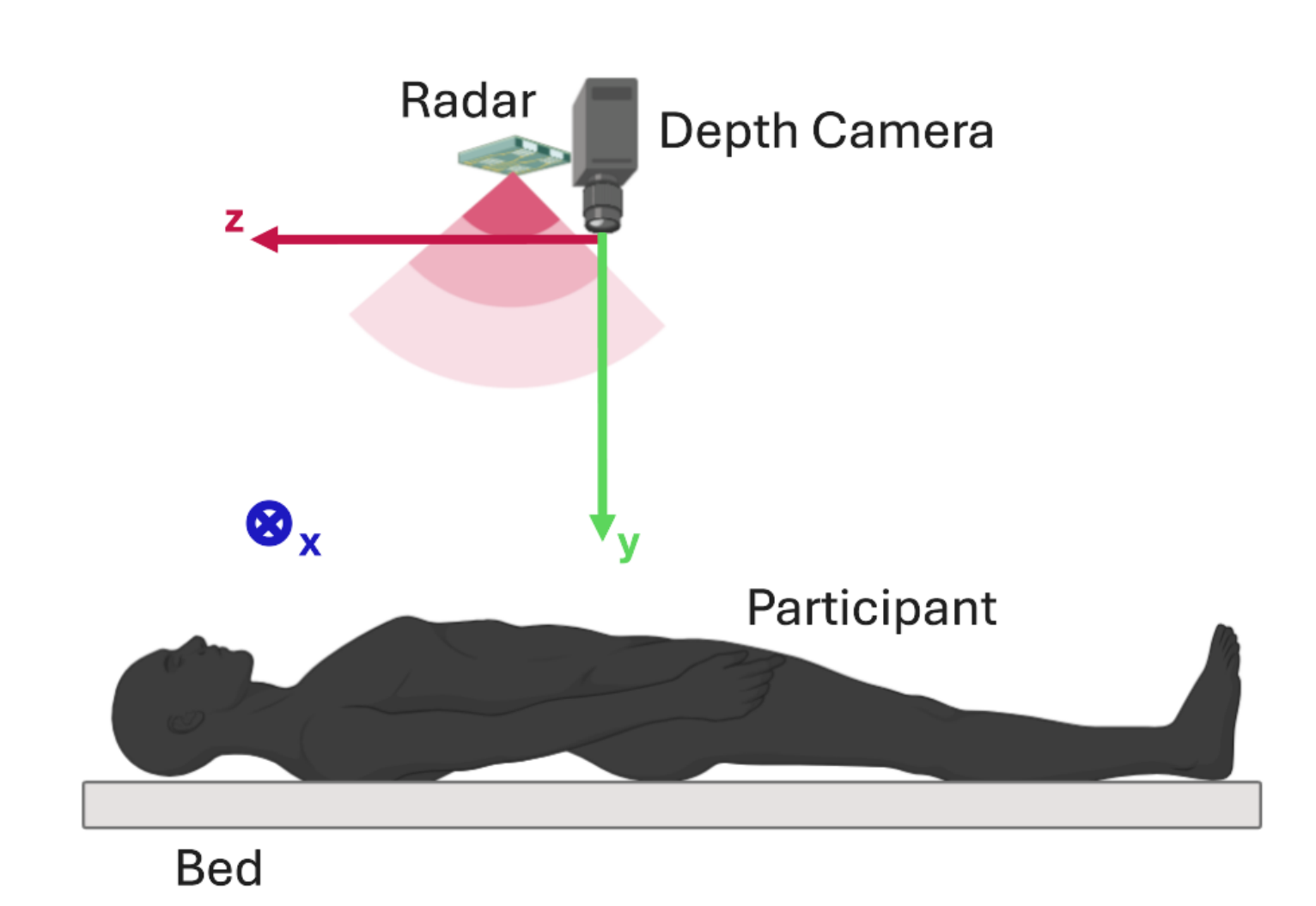}
            \subcaption{} \label{fig:schematic}
        \end{minipage}
        \caption{(a) Photograph and (b) schematic of the experimental setup, showing the depth camera, radar system, and a participant lying supine on a bed with their upper body exposed.}
        \label{fig:scene_schematic}
\end{figure}

A photograph and schematic illustration of the experimental setup are shown in Fig.~\ref{fig:scene_schematic}. The schematic also defines the Cartesian coordinate system adopted throughout the modeling pipeline. The origin of the coordinate system is placed at the optical center of the depth camera module, and the center of the radar array is located at
$x_{\mathrm{radar}} = 2.80 \times 10^{-2}\,\mathrm{m}$,
$y_{\mathrm{radar}} = -3.40 \times 10^{-2}\,\mathrm{m}$, and
$z_{\mathrm{radar}} = 1.38 \times 10^{-1}\,\mathrm{m}$.

\begin{table}[tb]
  \centering
  \caption{RADAR SYSTEM PARAMETERS (MODEL: T14RE\_01080108\_2D)}
  \label{tab:radar-spec}
  \begin{tabular}{ll}
    \toprule
    \textbf{Specification} & \textbf{Details} \\
    \midrule
    Operating principle & FMCW modulation \\
    Center frequency & 79.0 GHz \\
    Wavelength at center frequency & 3.8 mm \\
    Signal bandwidth & 3.6 GHz \\
    Number of Tx antennas & 3 \\
    Number of Rx antennas & 4 \\
    Tx antenna spacing & 7.6 mm \\
    Rx antenna spacing & 1.9 mm \\
    Tx beamwidth (E-plane / H-plane) & $\pm4^\circ$ / $\pm33^\circ$ \\
    Rx beamwidth (E-plane / H-plane) & $\pm4^\circ$ / $\pm45^\circ$ \\
    Achievable range resolution & 44.7 mm \\
    Sampling frequency (slow-time) & 100 Hz \\
    \bottomrule
  \end{tabular}
\end{table}

The spatial arrangement of the devices was determined on the basis of three primary considerations: avoiding mutual occlusion of the radar and depth camera fields of view, ensuring approximately uniform surface sampling by positioning the depth camera above the torso, and enabling accurate calibration by placing the two devices side by side above the torso, as shown in Fig.~\ref{fig:scene_schematic}.The resulting configuration is both intuitive and practical.

The vertical distance between the devices and the participant was set to approximately $0.8\,\mathrm{m}$. This distance represents a compromise between satisfying the far-field assumption required for electromagnetic scattering analysis and limiting the degradation of spatial resolution inherent to the depth camera at larger distances. While this distance was nominally fixed, slight variations occurred across participants due to differences in body shape and posture.

The experimental procedure was repeated for each participant as follows:
\begin{enumerate}
    \item The participant (adult male) lay supine on the bed with their upper body exposed.
    \item Respiratory motion of the chest and the abdomen was simultaneously recorded using the depth camera and the radar system.
    \item Immediately after the measurement, while maintaining the same posture, the upper body surface was scanned using the 3D scanner at a sampling resolution of $0.3\, \mathrm{mm}$ to obtain high-fidelity geometric data.
\end{enumerate}

The acquired data were processed using the proposed modeling framework described in Section~\ref{sec3}. Datasets from three participants were collected and used to validate the proposed approach by comparing the simulated IF signals with the corresponding radar measurements.

\subsection{Displacement Waveform Estimation}
\label{subsec4b}
To demonstrate the validity of the proposed modeling framework, we adopted a conventional and well-established signal processing pipeline for array radar as a baseline. Specifically, a range--angle map was constructed using FFT-based beamforming.

The range profile is derived by applying a Fourier transform to the IF signal defined in Eq.~\eqref{eq:IF_t} with respect to the fast-time variable $\tau$. By expressing the frequency axis $f$ in terms of range, $r = fc/(2\gamma)$, the resulting range-domain signal is given by
\begin{equation}
\begin{aligned}
            s_{\mathrm{R},i}(r, t) &= \sum_{k=1}^{K} |\bm{E}_\mathrm{scat}(\bm{r}_i;\bm{p}_k;t)|\eta_{i,k} \mathrm{sinc}\left\{\frac{4 B}{c} (r - R_{i,k}(t))\right\} \\
            &\quad \exp \left\{ \jj \left(\frac{f_\mathrm{min} R_{i,k}(t)}{c}\right) \right\},
\end{aligned}
\label{eq:range profile}
\end{equation}
where $\mathrm{sinc}(x) = \sin (\pi x) / (\pi x)$, and the scattering phase term $\eta_{i,k}$ is assumed to be a $\pi$-phase shift.

Beamforming is then applied across the linear virtual array with inter-element spacing $d_0$. For virtual channel indices $i = 0, 1, \ldots, N_\mathrm{R}-1$, the beamformed signal is expressed as
\begin{align}
  S(r, \theta, t) &= \sum_{i=0}^{N_\mathrm{R}-1} w_{i}(\theta) s_{\mathrm{R},i}(r, t) \nonumber \\
  &= \sum_{i=0}^{N_\mathrm{R}-1} \exp \left(-\jj \frac{2\pi}{\lambda_\mathrm{radar}}i d_0\sin\theta\right) s_{\mathrm{R},i}(r, t) .
\end{align}

Here, $\theta$ denotes the azimuth angle defined with respect to the Cartesian coordinate system shown in Fig.~\ref{fig:schematic}, and $\lambda_\mathrm{radar}$ is the wavelength corresponding to the radar center frequency. The resulting complex-valued radar image $S(r, \theta, t)$ is converted into a range--angle map by computing its power, $|S(r, \theta, t)|^2$.

The pixel exhibiting the maximum time-averaged power is identified as
\begin{equation}
(r_0, \theta_0) = \arg\max_{r, \theta} \left\{ \frac{1}{T} \sum_{t} \left| S(r, \theta, t) \right|^2 \right\},
\end{equation}
where $T$ denotes the total observation duration. The signal at this location, $S_{r_0,\theta_0}(t) = S(r_0, \theta_0, t)$, is then used to estimate the skin displacement $d(t)$ as
\begin{equation}
    d(t) = \frac{\lambda_\mathrm{radar}}{4 \pi} \mathrm{unwrap}\left(\angle S_{r_0,\theta_0}(t)\right) ,
\end{equation}
based on the approximate phase--displacement relationship $\angle S_{r_0,\theta_0}(t) \approx 4\pi d(t) / \lambda_\mathrm{radar}$.

In line with standard baseline processing, the estimated displacement waveform is further refined using smoothing and de-trending to suppress noise and highlight physiological motion. In addition, non-smoothed waveforms are also presented to illustrate the raw performance of the proposed modeling framework.

\subsection{Performance Evaluation}
\label{subsec4c}
\begin{figure*}[tb]
  \centering
  \renewcommand{\thesubfigure}{\alph{subfigure}}

  \begin{minipage}[t]{0.24\linewidth}
    \centering
    \includegraphics[height=4.5cm]{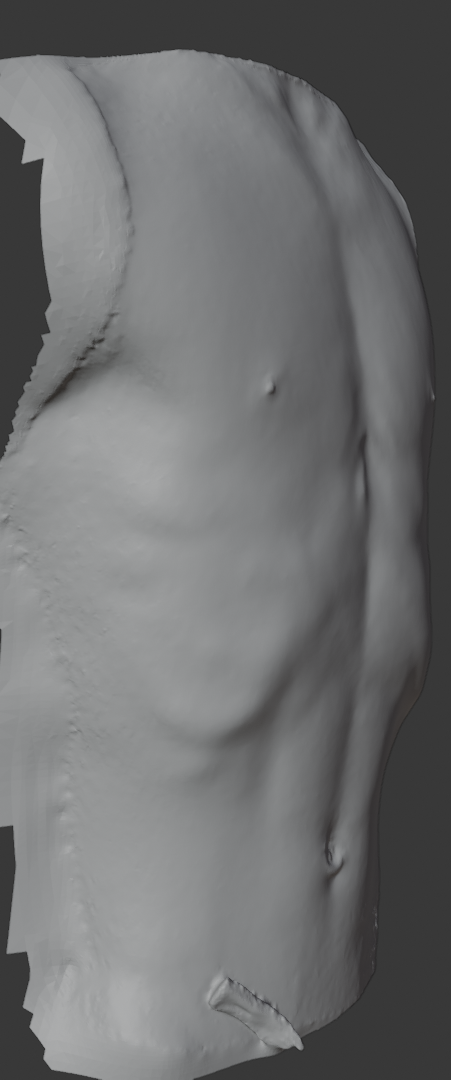}
    \subcaption{A: End-expiratory surface}\label{fig:a_exp_surface}
  \end{minipage}\hfill
  \begin{minipage}[t]{0.24\linewidth}
    \centering
    \includegraphics[height=4.5cm]{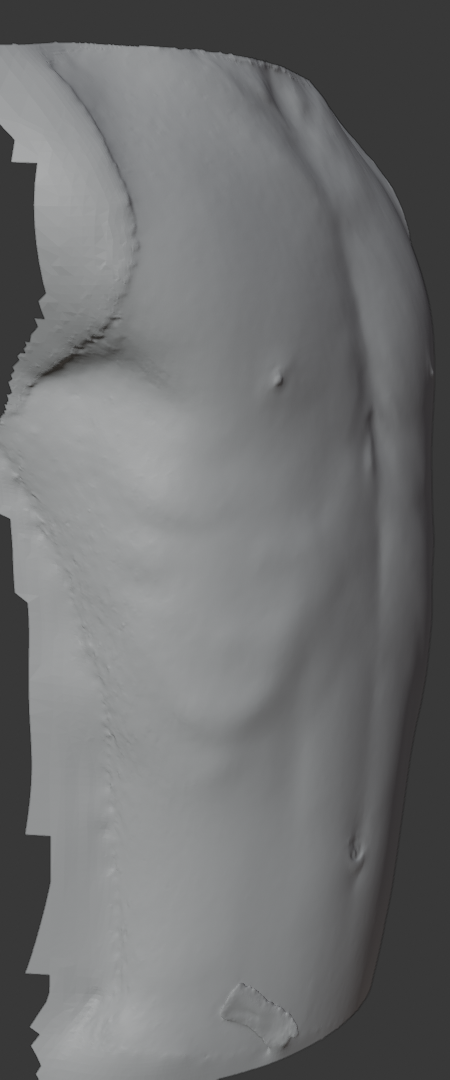}
    \subcaption{A: End-inspiratory surface}\label{fig:a_insp_surface}
  \end{minipage}\hfill
  \begin{minipage}[t]{0.24\linewidth}
    \centering
    \includegraphics[height=4.5cm]{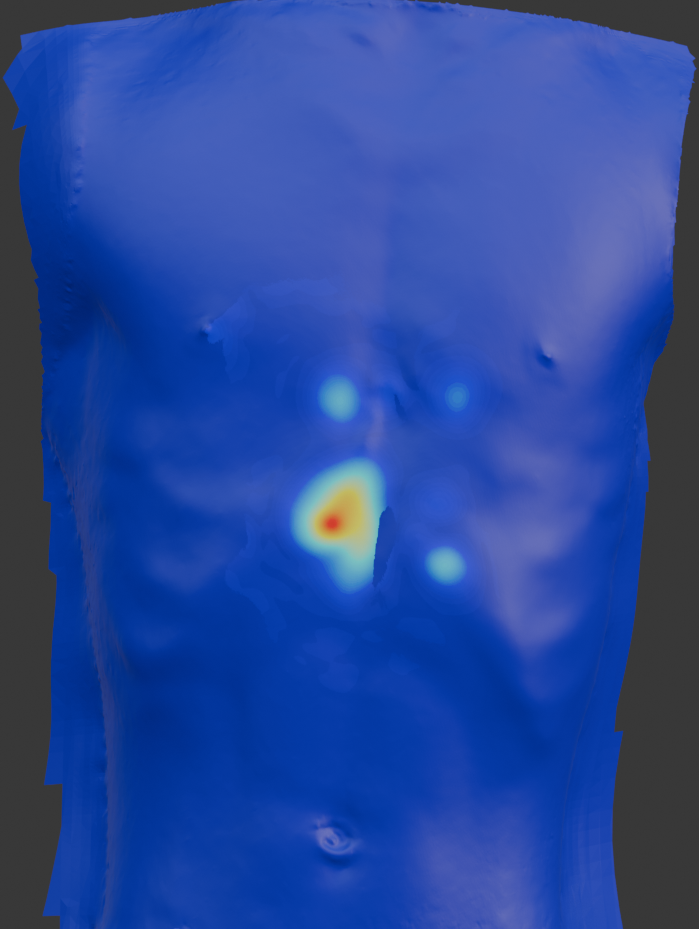}
    \subcaption{A: End-expiratory scattering}\label{fig:a_exp_scat}
  \end{minipage}\hfill
  \begin{minipage}[t]{0.24\linewidth}
    \centering
    \includegraphics[height=4.5cm]{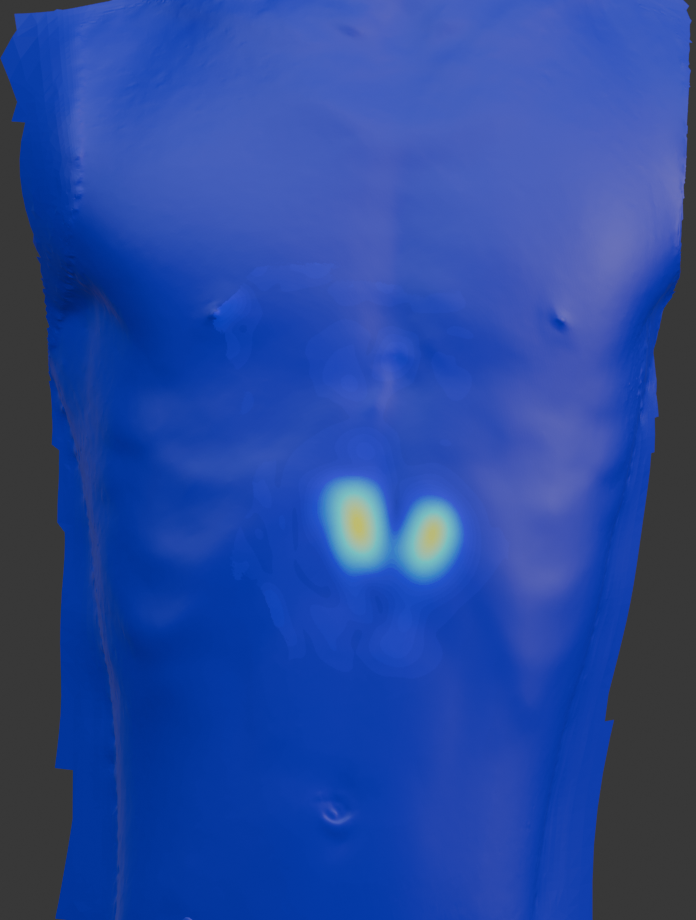}
    \subcaption{A: End-inspiratory scattering}\label{fig:a_insp_scat}
  \end{minipage}\hfill

  \vspace{1em}

  \begin{minipage}[t]{0.24\linewidth}
    \centering
    \includegraphics[height=4.5cm]{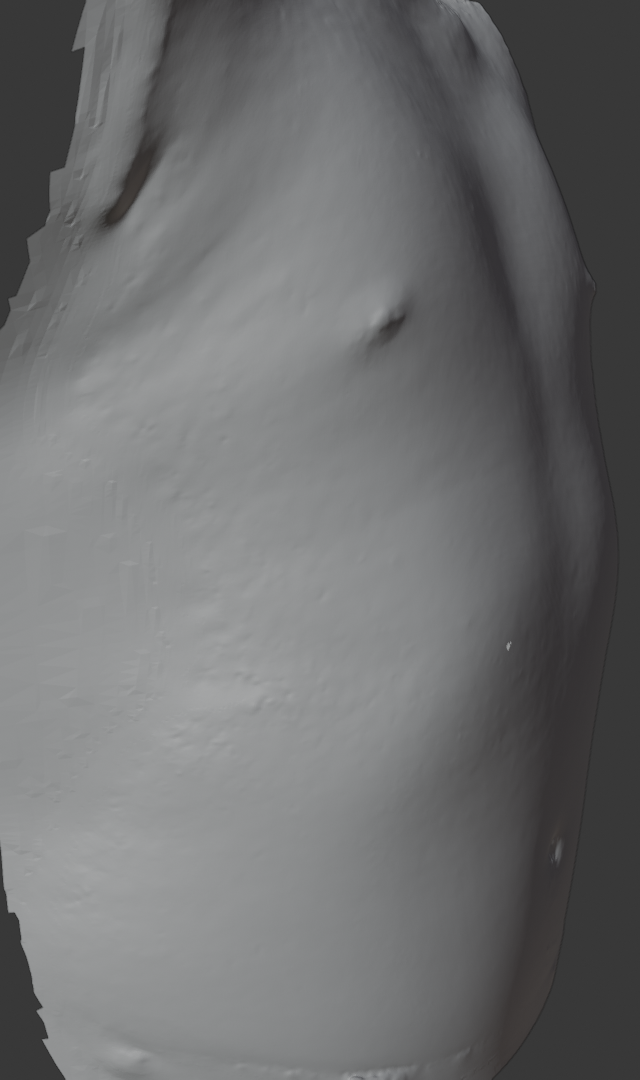}
    \subcaption{B: End-expiratory surface}\label{fig:b_exp_surface}
  \end{minipage}\hfill
  \begin{minipage}[t]{0.24\linewidth}
    \centering
    \includegraphics[height=4.5cm]{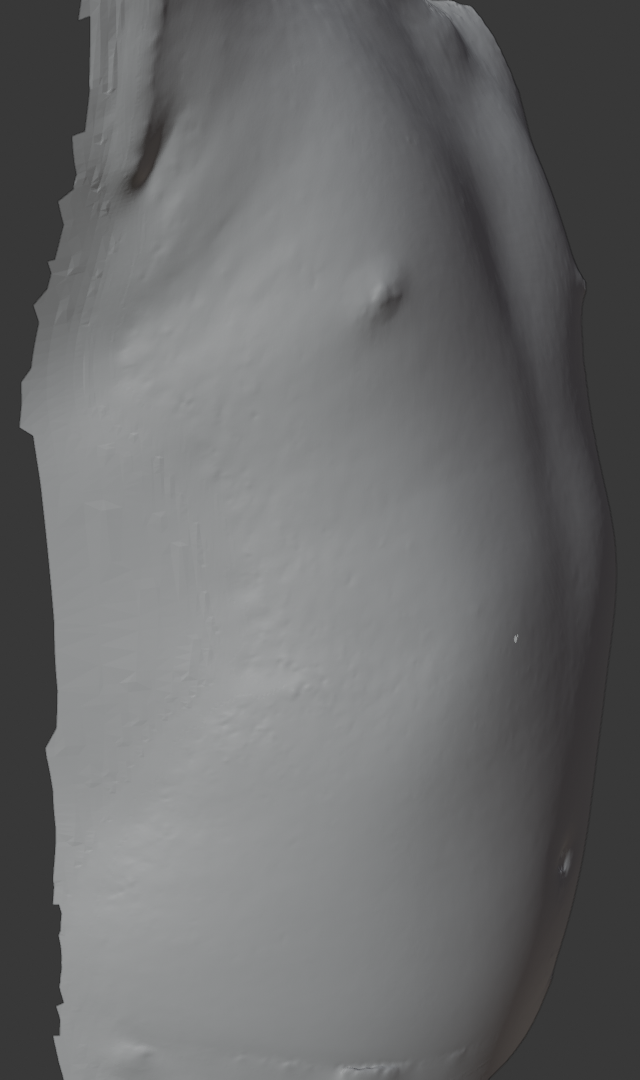}
    \subcaption{B: End-inspiratory surface}\label{fig:b_insp_surface}
  \end{minipage}\hfill
  \begin{minipage}[t]{0.24\linewidth}
    \centering
    \includegraphics[height=4.5cm]{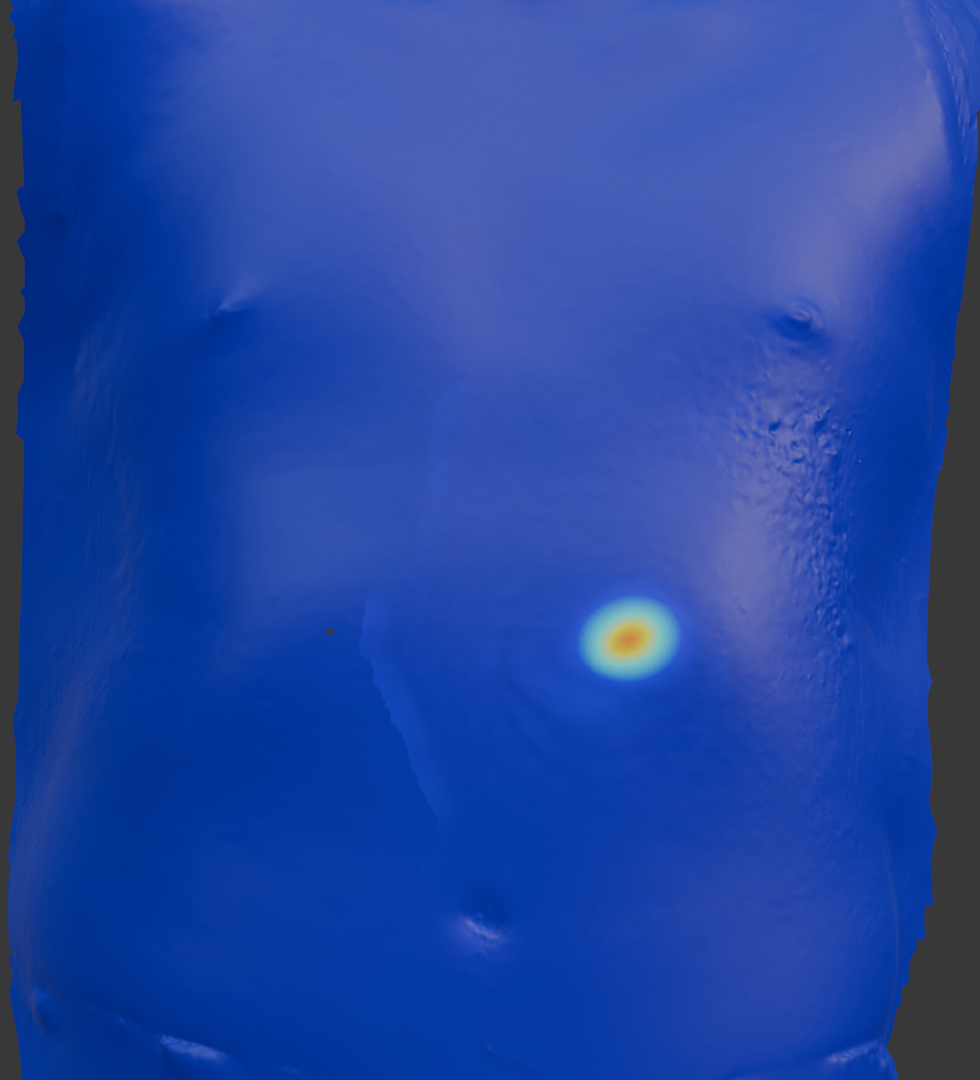}
    \subcaption{B: End-expiratory scattering}\label{fig:b_exp_scat}
  \end{minipage}\hfill
  \begin{minipage}[t]{0.24\linewidth}
    \centering
    \includegraphics[height=4.5cm]{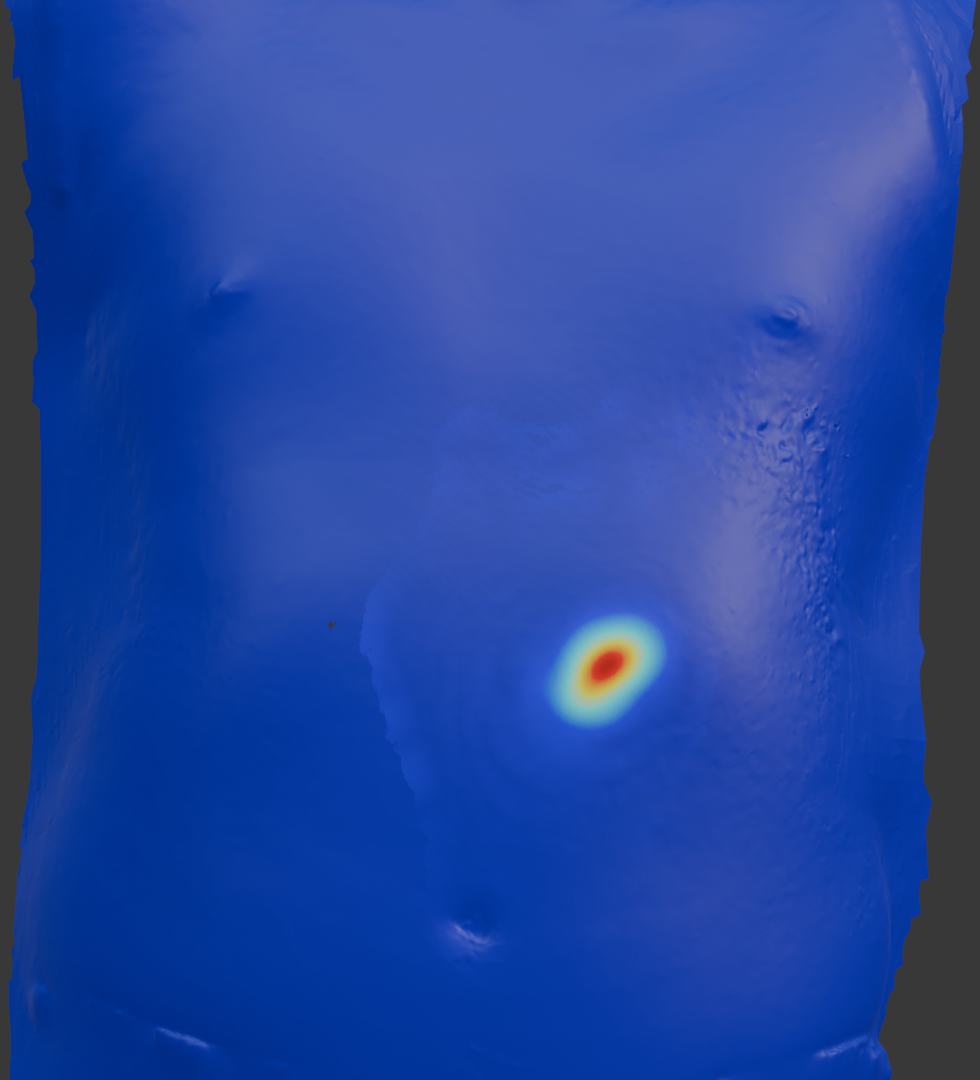}
    \subcaption{B: End-inspiratory scattering}\label{fig:b_insp_scat}
  \end{minipage}\hfill

  \vspace{1em}

  \begin{minipage}[t]{0.24\linewidth}
    \centering
    \includegraphics[height=4.5cm]{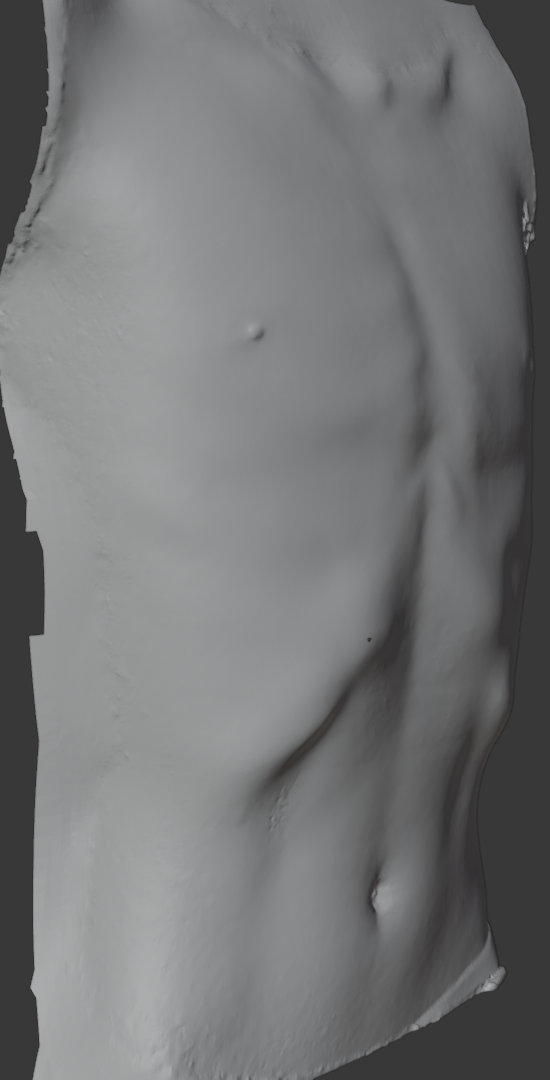}
    \subcaption{C: End-expiratory surface}\label{fig:c_exp_surface}
  \end{minipage}\hfill
  \begin{minipage}[t]{0.24\linewidth}
    \centering
    \includegraphics[height=4.5cm]{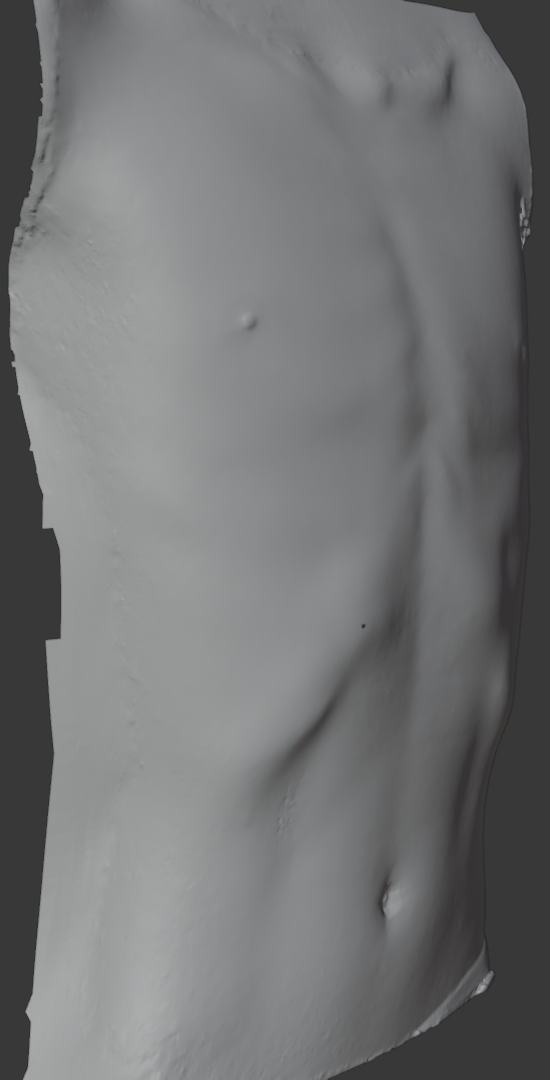}
    \subcaption{C: End-inspiratory surface}\label{fig:c_insp_surface}
  \end{minipage}\hfill
  \begin{minipage}[t]{0.24\linewidth}
    \centering
    \includegraphics[height=4.5cm]{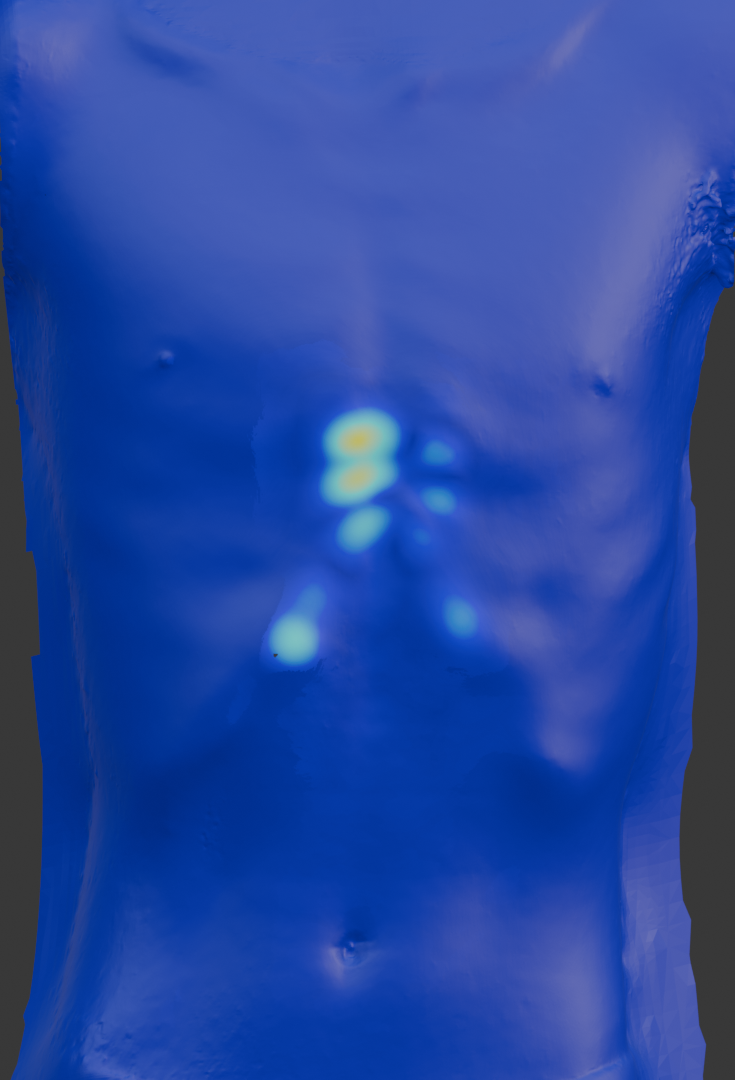}
    \subcaption{C: End-expiratory scattering}\label{fig:c_exp_scat}
  \end{minipage}\hfill
  \begin{minipage}[t]{0.24\linewidth}
    \centering
    \includegraphics[height=4.5cm]{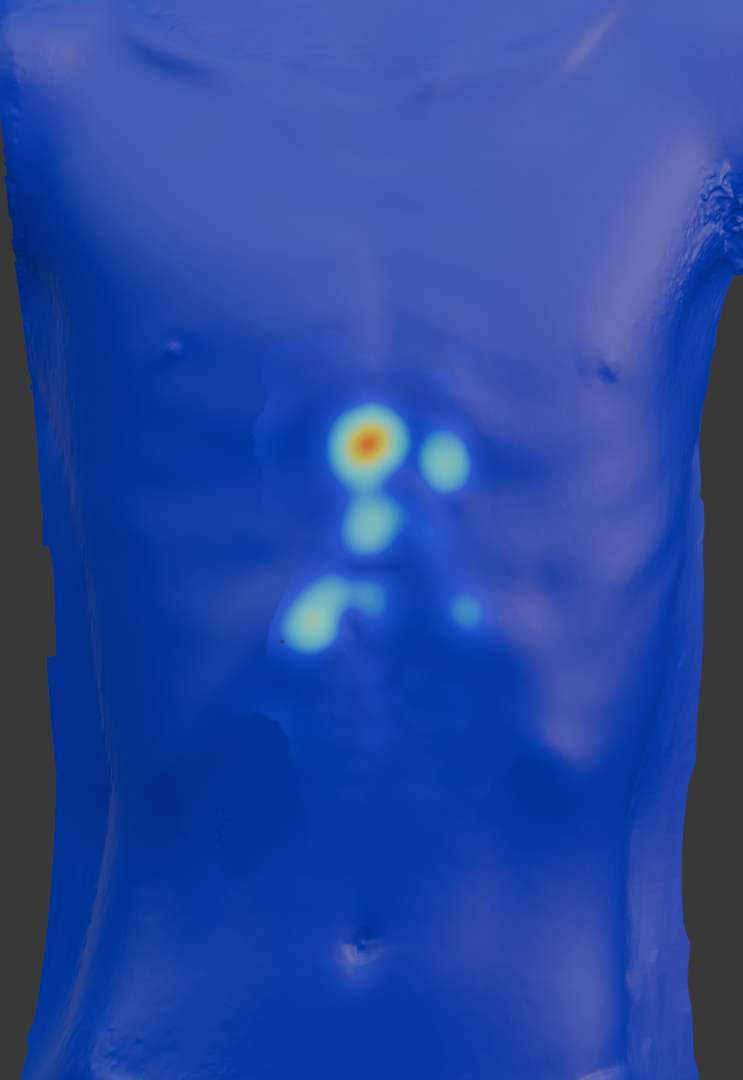}
    \subcaption{C: End-inspiratory scattering}\label{fig:c_insp_scat}
  \end{minipage}\hfill

  \caption{Reconstructed surface meshes and estimated scattering power distributions for each participant. For each subject (A, B, and C), the end-expiratory and end-inspiratory frames are shown side by side. Left columns show the reconstructed surface meshes and right columns show the corresponding scattering power distributions estimated using the PO-based analysis.}
  \label{fig:integration_scattering_all}
\end{figure*}

Data integration was performed for each participant by fitting the 3D scanner data to the depth camera sequences using the framework described in subsection~\ref{subsec3b}. Figure ~\ref{fig:integration_scattering_all} shows the reconstructed surface meshes and the corresponding estimated scattering power distributions at end-expiratory and end-inspiratory frames. Inter-participant differences in body geometry and respiratory motion patterns are clearly observable.

Participant A exhibits deeper breathing in which the dominant scattering regions (i.e., reflective sites) migrate along the sternum and costal joints toward the upper abdomen during inspiration. In contrast, participant B shows shallower breathing characterized by a single reflective site near the lower ribs that shifts downward toward the abdomen during inhalation. Participant C also demonstrates shallow breathing, but with multiple reflective sites distributed across the sternum and rib junctions that gradually converge as the rib cage expands.

Using the estimated scattering power distributions, we reconstructed IF signals following the procedures described above and subsequently extracted displacement waveforms. These waveforms were compared with the corresponding experimental radar measurements. For reference, a conventional modeling framework relying solely on depth camera data \cite{sumi} was also implemented.

\begin{figure*}[tb]
    \centering
    \begin{minipage}{0.32\linewidth}
        \centering
        \includegraphics[width=\linewidth]{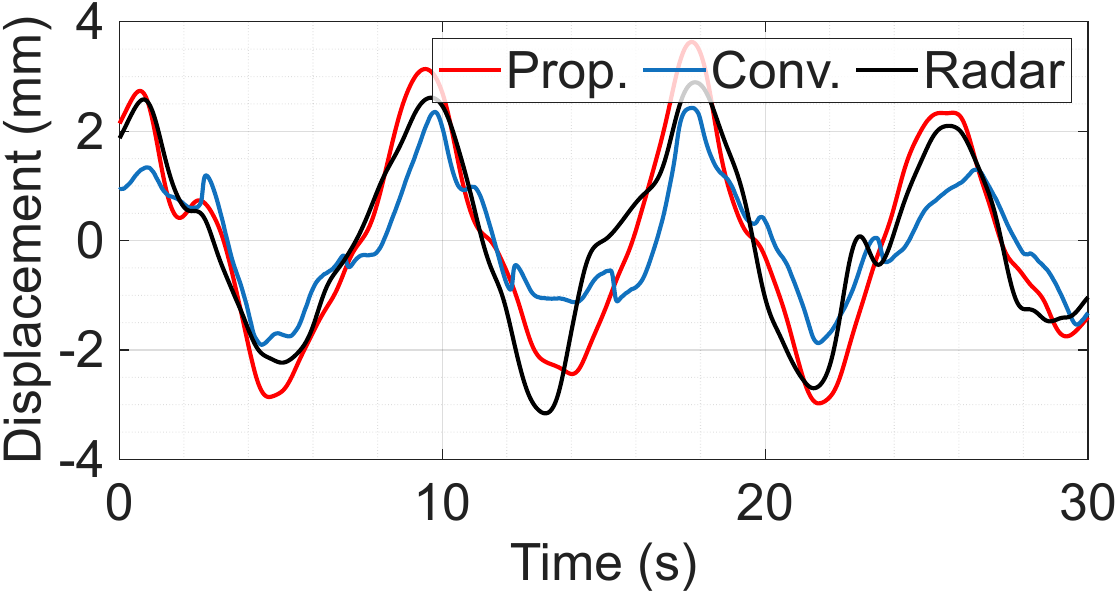}
        \subcaption{Participant A (Smoothed)} \label{fig:A_smoothed}
    \end{minipage}
    \hfill
    \begin{minipage}{0.32\linewidth}
        \centering
        \includegraphics[width=\linewidth]{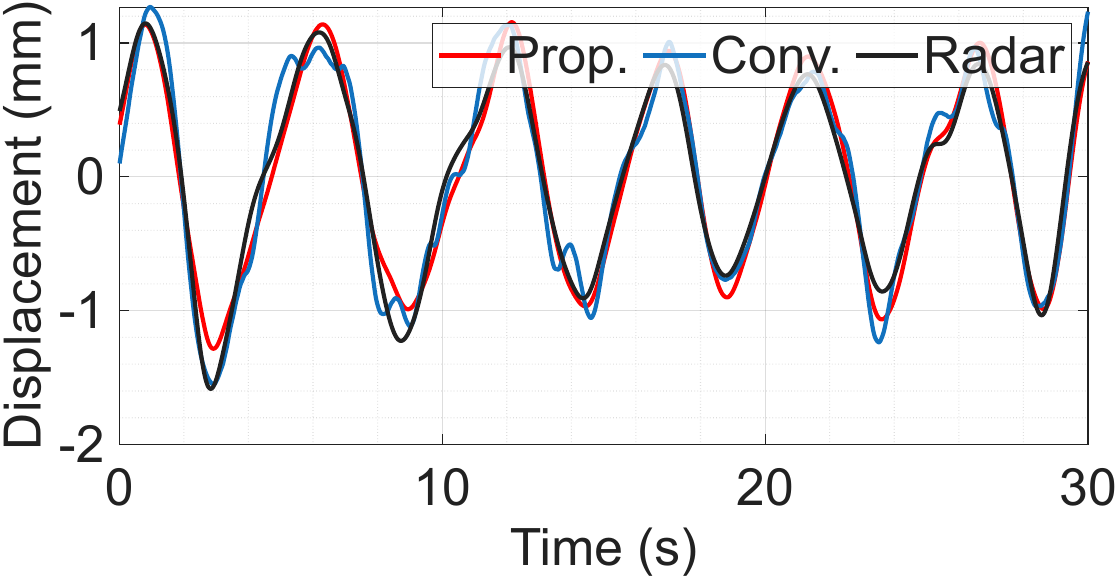}
        \subcaption{Participant B (Smoothed)} \label{fig:B_smoothed}
    \end{minipage}
    \hfill
    \begin{minipage}{0.32\linewidth}
        \centering
        \includegraphics[width=\linewidth]{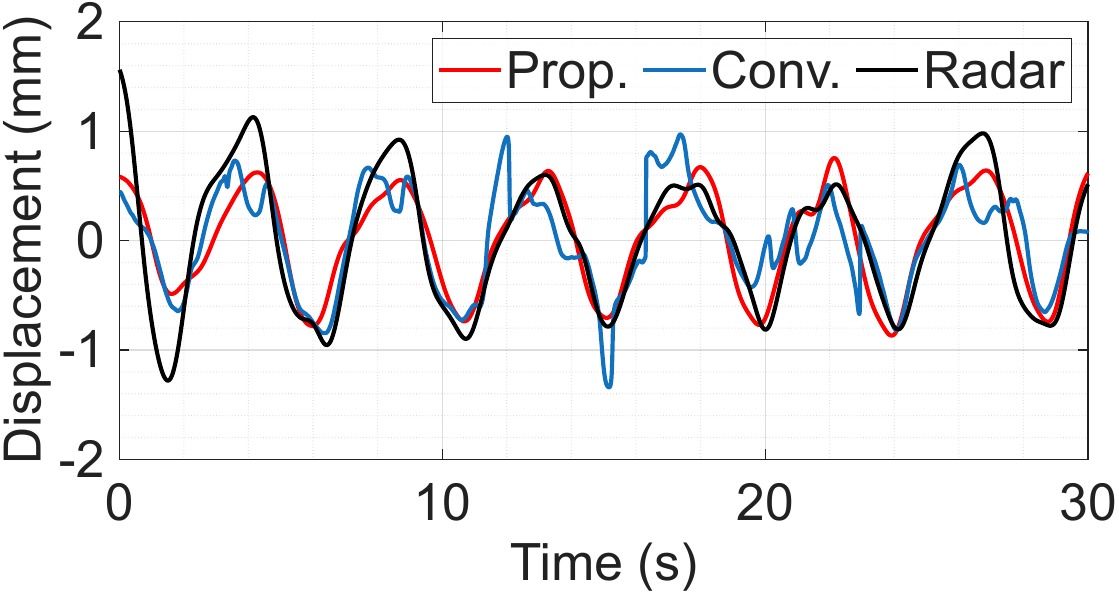}
        \subcaption{Participant C (Smoothed)} \label{fig:C_smoothed}
    \end{minipage}

    \vspace{1em}

    \begin{minipage}{0.32\linewidth}
        \centering
        \includegraphics[width=\linewidth]{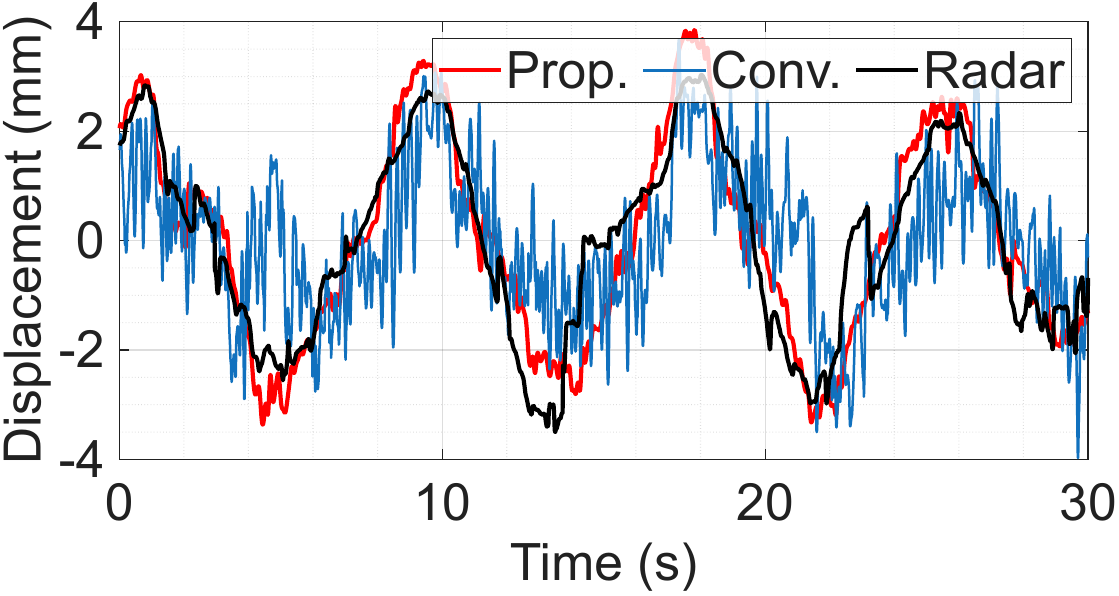}
        \subcaption{Participant A (Non-smoothed)} \label{fig:A_non-smoothed}
    \end{minipage}
    \hfill
    \begin{minipage}{0.32\linewidth}
        \centering
        \includegraphics[width=\linewidth]{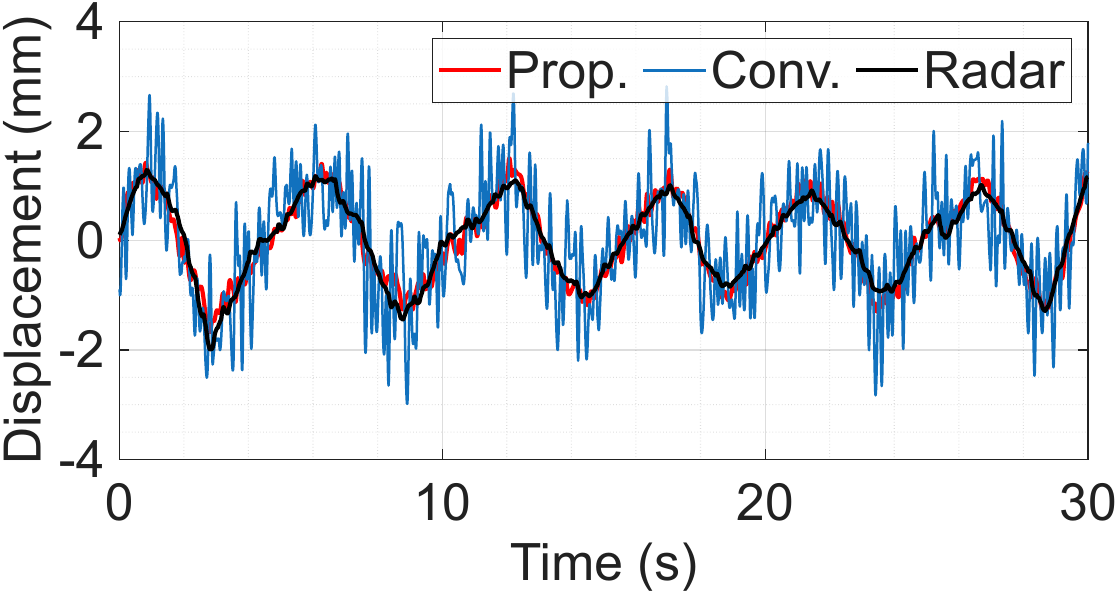}
        \subcaption{Participant B (Non-smoothed)} \label{fig:B_non-smoothed}
    \end{minipage}
    \hfill
    \begin{minipage}{0.32\linewidth}
        \centering
        \includegraphics[width=\linewidth]{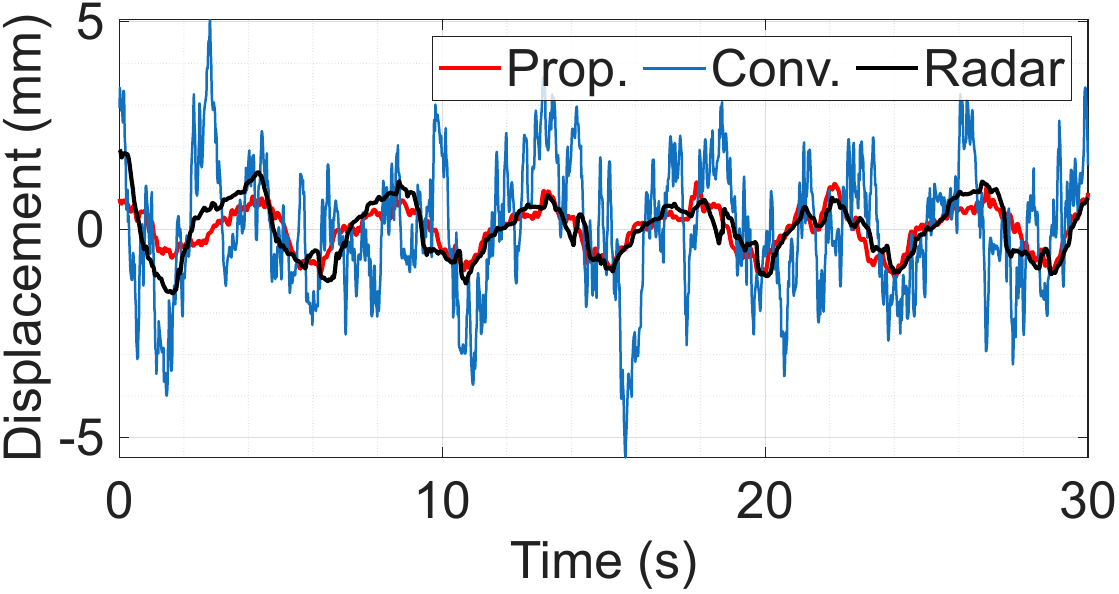}
        \subcaption{Participant C (Non-smoothed)} \label{fig:C_non-smoothed}
    \end{minipage}

    \caption{Displacement waveforms for all participants. Top row: refined waveforms obtained from the proposed modeling framework, the conventional modeling framework, and experimental radar data. Bottom row: non-smoothed waveforms obtained from the proposed modeling framework, the conventional modeling framework, and experimental radar data. The proposed method consistently demonstrates higher agreement with the experimental data across all participants.}
    \label{fig:disp_waveform_all}
\end{figure*}

Because the temporal synchronization between the depth camera and the radar system was imperfect, the maximum cross-correlation coefficient (Max Corr.) computed with time-shift compensation was adopted as the primary evaluation metric. In addition, the root-mean-squared (RMS) error and the Pearson correlation coefficient (PCC) between model-derived and experimentally measured displacement waveforms were calculated and reported as secondary performance indicators. As shown in Fig.~\ref{fig:disp_waveform_all}, the proposed modeling framework exhibits consistently higher agreement with the experimental radar measurements across all participants than that of the depth-sequence-only conventional model.

Quantitative results supporting these observations are summarized in Table~\ref{tab:eval_smooth_nosmooth}, which reports RMS error, Max Corr., and a PCC for both the conventional and proposed methods. The proposed approach consistently outperforms the conventional approach across all evaluation metrics for all three participants.

The same metrics were also used to evaluate the non-smoothed waveforms generated by the proposed modeling framework, with these evaluations demonstrating that its performance does not rely on smoothing. In contrast, the conventional modeling framework incorporates smoothing as an integral part of its IF signal reconstruction process, and therefore exhibits strong dependence on the smoothing procedure.

\begin{table}[tb]
  \caption{Evaluation metrics with and without smoothing for each participant. Root-mean-squared (RMS) error, maximum cross-correlation coefficient (Max Corr.), and Pearson correlation coefficient (PCC) between model-derived and experimentally measured displacement waveforms are reported for the conventional (Conv.) and proposed (Prop.) methods.}

  \label{tab:eval_smooth_nosmooth}
  \centering
  \renewcommand{\arraystretch}{1.2}
  \resizebox{\linewidth}{!}{%
    \begin{tabular}{@{}clccc@{}}
    \toprule
    \textbf{Condition} & \textbf{Participant} & \textbf{RMS Error (mm)} & \textbf{Max Corr.} & \textbf{PCC} \\
    \midrule

    \multirow{6}{*}{\textbf{Smoothed}}
    & A (Conv.)  & 0.892 & 0.904 & 0.868 \\
    & A (Prop.)  & 0.599 & 0.944 & 0.943 \\
    \cmidrule{2-5}
    & B (Conv.)  & 0.167 & 0.975 & 0.975 \\
    & B (Prop.)  & 0.121 & 0.985 & 0.985 \\
    \cmidrule{2-5}
    & C (Conv.)  & 0.365 & 0.796 & 0.796 \\
    & C (Prop.)  & 0.288 & 0.887 & 0.887 \\
    \midrule

    \multirow{6}{*}{\textbf{No Smoothing}}
    & A (Conv.)  & 1.42 & 0.669 & 0.587 \\
    & A (Prop.)  & 0.712 & 0.925 & 0.924 \\
    \cmidrule{2-5}
    & B (Conv.)  & 0.665 & 0.753 & 0.752 \\
    & B (Prop.)  & 0.179 & 0.971 & 0.971 \\
    \cmidrule{2-5}
    & C (Conv.)  & 1.41 & 0.442 & 0.442 \\
    & C (Prop.)  & 0.388 & 0.824 & 0.824 \\
    \bottomrule
  \end{tabular}%
}
\end{table}

\begin{figure}[tb]
    \centering
    \includegraphics[width=\linewidth]{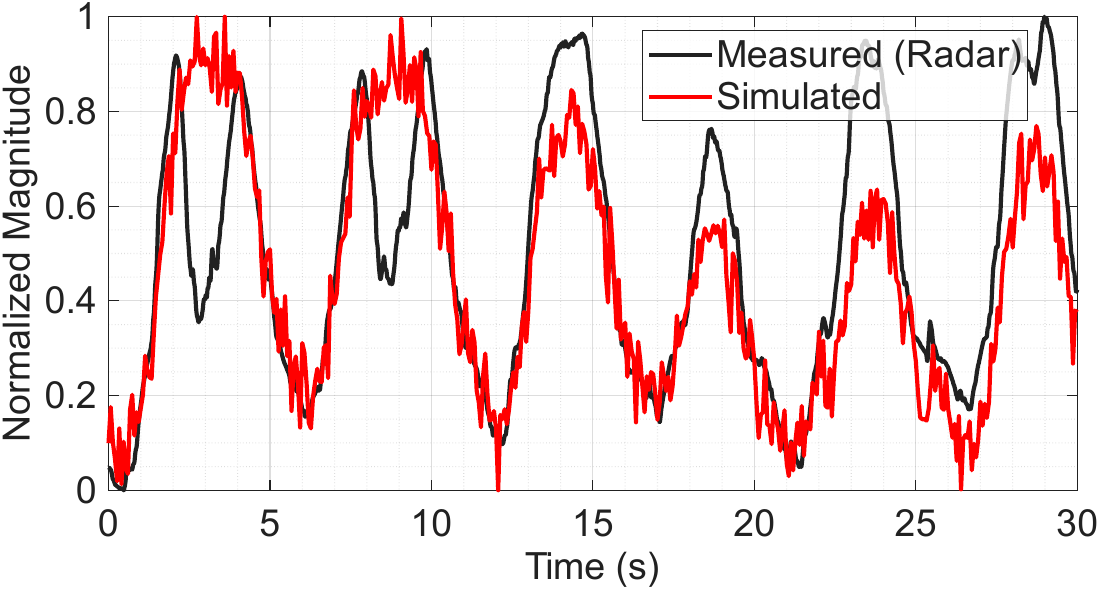}
    \caption{Magnitude of the I--Q signal over time for Participant B, illustrating agreement between simulated and experimental radar signals in a representative high-quality case.}
    \label{fig:participant_B_mag}
\end{figure}

The radar data acquired from Participant B exhibited significantly higher signal quality than the data from Participants A and C. Figure~\ref{fig:participant_B_mag} presents a follow-up analysis of this high-quality case, which shows that the simulated in-phase and quadrature (I--Q) signal, whose phase is conventionally used for displacement estimation, demonstrates strong magnitude agreement with the corresponding experimental radar signal. In this case, a PCC of $0.789$ between model-derived and experimentally measured I--Q magnitude was obtained at a compensated time lag of $0.07$ s, corresponding to the time shift that maximized the cross-correlation of the raw displacement waveform.

This level of agreement was observed only for participant B, for whom a single reflective site remained relatively stationary throughout the respiratory cycle. In contrast, Participants A and C represent lower signal quality scenarios characterized by multiple reflective sites that shift during respiration. Under such conditions, the baseline displacement estimation approach struggles to produce satisfactory results. These observations indicate that the conventional modeling framework is more susceptible to degraded performance in the presence of complex surface dynamics, whereas the proposed framework demonstrates greater robustness across the tested cases.

\section{Conclusion}
\label{sec:conclusion}
This study proposes a deformation-aware observation modeling approach  for radar-based respiratory sensing that integrates high-resolution 3D scanner data with time-series depth camera measurements to capture dynamic surface deformation of the human torso during respiration. A static 3D-scan-derived surface representation is first extracted as a geometric template. A CPD-based non-rigid registration procedure is then employed to fit the template to each depth frame, yielding a time-varying deformation model of the torso. Based on this dynamic surface representation, frame-wise electromagnetic scattering is estimated using a PO approximation, and radar IF signals are reconstructed to simulate radar observations.

Experimental validation demonstrated that the proposed framework consistently achieved higher maximum cross-correlation coefficients than a conventional modeling approach relying solely on depth camera data. In a high-quality case characterized by a single relatively-stationary reflective site, the proposed framework reproduced I--Q magnitude variations with a PCC of 0.789, indicating its ability to accurately capture radar measurement characteristics under favorable scattering conditions. Furthermore, in lower-signal-quality scenarios involving complex surface dynamics and multiple reflective sites, the proposed framework exhibited improved robustness and maintained superior agreement with experimental radar measurements, achieving higher cross-correlation values than the conventional depth-sequence-only approach.

Several practical limitations should be acknowledged. First, no temporal filtering or smoothing is applied to the reconstructed mesh sequence, which may lead to frame-to-frame discontinuities, particularly in sparsely sampled peripheral regions. Second, although the CPD-based registration mitigates spatial randomness in depth camera measurements, residual noise may still propagate into the scattering estimates and reconstructed IF signals. Third, the current implementation remains computationally demanding, with processing times on the order of several hours per participant, despite the use of cost-reduction strategies. Future work will focus on addressing these limitations by improving temporal coherence, enhancing noise robustness, and reducing computational cost, thereby working towards a more practical and physically grounded deformation-aware radar observation model for radar-based respiratory measurement.

\section*{Acknowledgment}
\addcontentsline{toc}{section}{Acknowledgment}
\scriptsize
This work was supported in part by JST SPRING Grant JPMJSP2110; the SECOM Science and Technology Foundation; in part by Japan Science and Technology Agency under Grant JPMJMI22J2 and Grant JPMJMS2296; in part by the Japan Society for the Promotion of Science KAKENHI under Grant 21H03427, Grant 23H01420, and Grant 23K26115; and in part by the New Energy and Industrial Technology Development Organization.
\normalsize

\section*{Ethics Declarations}
The experimental protocol involving human participants was approved by the Ethics Committee of the Graduate School of Engineering, Kyoto University (permit no. 202223). Informed consent was obtained from all human participants in this study.

\end{document}